\documentclass[useAMS,usenatbib,usegraphicx]{mn2e}

\def\phn{\phantom{0}}       

\def\lesssim{\mathrel{\hbox{\rlap{\hbox{\lower4pt\hbox{$\sim$}}}\hbox{$<$}}}}
\def\gtrsim{\mathrel{\hbox{\rlap{\hbox{\lower4pt\hbox{$\sim$}}}\hbox{$>$}}}}

\newcommand{\unit}[1]{\ifmmode \:\mbox{\rm #1}\else \mbox{#1}\fi}
\newcommand{\sbr}[1]{_{\rm #1}}
\newcommand{\secref}[1]{Section~\ref{sec:#1}}
\newcommand{\appref}[1]{Appendix~\ref{sec:#1}}

\newcommand{\eqref}[1]{equation~(\ref{eq:#1})}

\newcommand{\figref}[1]{Fig.~\ref{fig:#1}}
\newcommand{\tabref}[1]{Table~\ref{tab:#1}}

\title[Probing Satellite Halos with Weak Gravitational Lensing]{Probing Satellite Halos with Weak Gravitational Lensing}
\author[Bryan R. Gillis {\it et al.}]{Bryan R. Gillis$^{1}$\thanks{E-mail:
bgillis@uwaterloo.ca}, Michael J. Hudson$^{1,2}$, Stefan Hilbert$^{3,4}$, Jan Hartlap$^{4}$ \\
$^{1}$Department of Physics and Astronomy, University of Waterloo, Waterloo, ON N2L 3G1, Canada.\\
$^{2}$Perimeter Institute for Theoretical Physics, 31 Caroline St. N., Waterloo, ON, N2L 2Y5, Canada.\\
$^{3}$Kavli Institute of Particle Astrophysics and Cosmology (KIPAC), Stanford University, 452 Lomita Mall, Stanford, CA 94305, and \\
SLAC National Accelerator Laboratory, 2575 Sand Hill Road, M/S 29, Menlo Park, CA 94025, United States of America.\\
$^{4}$Argelander-Institut f{\"u}r Astronomie, Universit{\"a}t Bonn, Auf dem H{\"u}gel 71, 53121 Bonn, Germany.}
\begin{document}

\date{??}

\pagerange{\pageref{firstpage}--\pageref{lastpage}} \pubyear{2011}

\maketitle

\label{firstpage}

\begin{abstract}
We demonstrate the possibility of detecting tidal stripping of dark matter subhalos within galaxy groups using weak gravitational lensing. We have run ray-tracing simulations on galaxy catalogues from the Millennium Simulation to generate mock shape catalogues. The ray-tracing catalogues assume a halo model for galaxies and groups, using various models with different distributions of mass between galaxy and group halos to simulate different stages of group evolution. Using these mock catalogues, we forecast the lensing signals that will be detected around galaxy groups and satellite galaxies, as well as test two different methods for isolating the satellites' lensing signals. A key challenge is to determine the accuracy to which group centres can be identified. We show that with current and ongoing surveys, it will possible to detect stripping in groups of mass $10^{12}$~$M_{\odot}$--$10^{15}$~$M_{\odot}$.
\end{abstract}


\section{Introduction}

\label{sec:intro}
In the standard picture of hierarchical structure formation, larger dark matter halos are built up through the accretion, stripping, and mergers of smaller halos. At the extremes of the halo mass spectrum, namely isolated field galaxies and galaxy clusters, we have a relatively good picture of how the mass within these structures is organized. For isolated galaxies, most of the mass is contained within a halo of dark matter, as confirmed by galaxy-galaxy weak gravitational lensing measurements \citep{BraBlaSma96, HudGwyDah98, SmiBerFis01, GuzSel02, ManSelKau06, GavTreRho07}. Simulations have shown that the shape of this halo can be well-approximated by an NFW density profile \citep{NavFreWhi97,LuMoKat06} and galaxy-galaxy weak gravitational lensing measurements have confirmed these results \citep{KleSchErb03,ManSelHir08}. In galaxy clusters, most of the mass also seems to lie within an NFW dark matter halo, with the constituent galaxies contributing only small perturbations to the density profile \citep{ManSelCoo06,ManSelHir08}. Gravitational lensing measurements have confirmed that the halos around individual galaxies within clusters are significantly smaller than the halos around comparably-luminous field galaxies, and this effect is more extreme with galaxies closer to the centres of clusters \citep{LimKneBar07,NatKneSma09}.

However, between the extremes of field galaxies and rich clusters, the picture is less clear. Since multiple galaxies must merge together to eventually form clusters, at some point the mass in the galaxies' individual halos must migrate into a shared halo. This process most likely occurs through tidal stripping: when two galaxies pass near each other, the halo of the less massive galaxy will tend to be ``stripped'' from it and thus join the more massive galaxy's halo. This effect has been demonstrated in various N-body dark matter simulations \citep{HayNavPow04,KazMayMas04,SprWanVog08}. Tidal stripping is also expected to remove hot gas from less massive galaxies, which will have the effect of quenching their star formation in a process known as ``strangulation'' \citep{BalMor00}. Galaxies in dense environments are known to be significantly redder on average than field galaxies \citep{ButOem84,MooKatLak96,BalMorYee99,BalBalNic04}, and tidal stripping may contribute to the quenching of star formation, so there is a strong motivation to understand the mechanics and timing of tidal stripping \citep{AquYan08,KawMul08}. It remains unclear, however, whether this process is rapid or gradual.

In this paper, we focus on galaxy groups, an intermediate mass scale between field galaxies and clusters (typically structures in the mass range ($10^{12}$~M$_{\odot} \lesssim M\sbr{halo} \lesssim 10^{14}$~M$_{\odot}$ are considered groups, and more massive structures are considered clusters). Weak gravitational lensing provides the only practical tool to measure the shapes and masses of dark matter halos around individual galaxies within groups.  Lensing analyses of groups \citep{HoeFraKui01, ParHudCar05, ManSelCoo06, JohSheWec07, HamMiyKas09, LeaFinKne10, ForHilVan12} have shown that the group lensing signal can be measured and is consistent with an NFW density profile. However, as we show in \secref{SigCen}, the signal could result from either a single, smooth halo, or a model which includes both a central halo and an individual halo for each satellite. As such, it is necessary to measure at the lensing signals around satellites themselves to get a full picture of the mass distribution. Only limited work has been done in the group regime to date. For example, \citet{SuyHal10} studied a strong-lensing system and determined that tidal stripping did seem to occur around the satellite studied, which lies in a group of mass on the order of $10^{12} M_{\odot}$. While this result is promising, a broader base of data will be needed to develop firm conclusions. 

Upcoming data from surveys such as the GAMA-II survey \citep{DriHilKel11} will provide an unprecedented opportunity for us to use weak gravitational lensing to analyze a large sample of galaxy groups. In anticipation of the availability of this data, in this paper we investigate various methods for detecting the extent of tidal stripping of satellite halos. We hope to determine the best analysis method and to generate a framework through which we can interpret future results, and assess the effects of various sources of bias and error. Previous theoretical studies \citep{MNatKne02, NatKneSma02, YanMo06} have looked into this issue. In particular, \citet{PasHilHar11} demonstrated  a method  -- called ``Galaxy-Galaxy Lensing with Calibration'' (GGLC) by them and the ``Mirror Method'' by us (see \secref{Mirror}) -- that can be used to isolate the lensing signals of satellite galaxies. Our work expands on these past results by testing and comparing different methods for isolating satellites' lensing signals from the contamination of group halos, assessing the effects of errors in estimating the position and mass of the group's halo on these results. To do so, we simulate the effects of weak gravitational lensing on a sample of mock galaxy catalogues extracted from the Millenium Simulation, using models that correspond to different amounts of tidal stripping. We measure the lensing signal around both satellites and group centres, and apply two different methods to isolate the lensing signal of the satellites' halos by removing the estimated signal contribution of the groups' halos. By looking at the resultant lensing signals for satellites of similar mass, but varying the stripping model, the mass of the groups that contain these satellites, and the magnitudes of sources of error, we can determine which method is superior, and in which regimes it might be possible to distinguish between stripping models.

It is possible to study galaxies in a group environment using either photometric or spectroscopic redshift estimates. Since photometric redshift estimates are significantly worse than spectroscopic estimates, with typical errors of $\sim 0.05$ in z for 5-band surveyrs \citep{Ben00,HilErbKui12}, it is thus much more difficult to generate a group catalogue of galaxies. Various methods exist to determine the positions of groups using only photometric redshifts \citep{KoeMcKAnn07,LiYee08,MilHey10,GilHud11}, but it is difficult to determine which galaxies belong to these groups and which are field galaxies \citep{GilHud11}. Spectroscopic redshifts provide the advantage that group membership can typically be determined to $\sim80\%$ purity and completeness \citep{RobNorDri11}. Although it is possible to investigate the presence or absence of tidal stripping using only photometric redshift estimates, these methods are beyond the scope of this paper, which deals only with the case where spectroscopic redshifts are available. Methods based only on photometric redshifts will be discussed in a future paper.

In \secref{data} of this paper, we discuss the galaxy catalogues we use and our models for halo mass distribution. In \secref{simulations}, we explain how we simulate lensing signals and how we remove the influence of groups' signals from the lensing signals detected around satellites. In \secref{analysis}, we present the results of our simulations and discuss their implications.

For consistency with the Millenium Simulation, we use the following cosmological parameters: $H\sbr{0} = 73$ km s$^{-1}$ Mpc$^{-1}$, $\Omega\sbr{m} = 0.25$, $\Omega\sbr{\lambda} = 0.75$, and $\Omega\sbr{b}=0.045$. All stated magnitudes are in the AB system. Since there is no clear division between galaxy groups and galaxy clusters, we use the terminology ``galaxy groups'' throughout this paper, even though some of the structures we refer to as such would be more commonly deemed clusters.  When masses are quoted in this paper, $M$ is used to refer to the halo mass of a group, and $m$ is used to refer to the halo mass of a satellite, unless otherwise specified. When radial measurements are used in this paper, $R$ refers to a projected, 2D distance, and $r$ refers to a 3D distance. 

\section{Simulations and Models}

In this section we discuss the simulations used to generate our galaxy catalogues and our models for the mass distributions of galaxies within these catalogues. In \secref{SimulatedData}, we discuss how we extracted our galaxy catalogues from the Millennium Simulation and the properties of these catalogues. In \secref{Models} we discuss the mass distributions we assume for dark matter halos surrounding the galaxies in our catalogues; \secref{halomodels} discusses the mathematical form of the model we use, and \secref{strmodels} discusses the different models we use for the presence or absence of tidal stripping, along with other datasets we use for comparison purposes.

\label{sec:data}
\label{sec:Data}

\subsection{Simulations}
\label{sec:SimulatedData}

In order to assess the differences varying stripping models have between their lensing signals, it is easiest to work with a simulated galaxy catalogue, where exact redshifts, masses, and mass distributions can be known. In this paper, we used a semi-analytic galaxy catalogue based on the Millennium Simulation \citep{SprWhiJen05} by \citet{DeBla07}. The Millennium Simulation is a collisionless simulation of $N = 2160^3$ dark matter particles, each with a mass of $8.6 \times 10^8\;h^{-1}$M$_{\odot}$, in a box with $500\;h^{-1}$Mpc sides and periodic boundary conditions. The simulation used the following cosmological parameters: $h = 0.73$, $\Omega_m = 0.25$, $\Omega_v = 0.75$, $\Omega_b = 0.045$, $\sigma_8 = 0.9$, $n = 1$. The simulation used force softening on a scale of $5\;h^{-1}$ comoving kpc. This causes an artificial smoothing of the cores of dark matter halos, and as a result, density measurements within $\sim50$ kpc of a halo's core are unreliable.

The galaxy catalogues used in this paper consist of lightcones within \citet{DeBla07}'s catalogue prepared by \citet{HilHarWhi09}. These catalogues are complete for $M\sbr{stellar} > 10^{9}$~M$_{\odot}$ and consist of thirty-two 16~deg.$^2$ fields. We split these catalogues into sub-catalogues of ``lens'' and ``source'' galaxies. The lens catalogues consist of all galaxies in the parent catalogue with $0.05 < z < 0.8$, and the source catalogues consist of all galaxies in the parent catalogue with $r < 24.5$, approximately the limit for shape data in the CHFLenS survey \citep{HeyMil12}. We use all galaxies in the lens catalogue for our ray-tracing simulations, but only those galaxies with $r < 19.8$ for analysis, matching the completeness limit for the GAMA-II survey \citep{DriHilKel11}. The source galaxies have shears calculated from a ray-tracing simulation performed by \citeauthor{HilHarWhi09}, which differs from ours in that it uses the positions of dark matter particles within the Millennium Simulation rather than our halo models (described in \secref{Models}). As such, it provides a useful comparison to check that our models generate reasonable results.

The resolution of redshift and mass within these catalogues was limited by the number of snapshots of the original simulation which were saved. Sixty-four snapshots were saved in total, of which 23 are below redshift 1.

The dark matter halos of the Millennium Simulation were populated with galaxies through a semi-analytic model developed by \citet{DeBla07}. Dark matter halos are identified through a friends-of-friends algorithm, and sufficiently massive halos are populated with galaxies. Halos are also analyzed with the SUBFIND algorithm \citep{SprWhiTor01} to identify subhalos, and self-bound subhalos are similarly populated with galaxies. All of these galaxies are labeled with the ID of the dark matter halo or subhalo that contains them. Halos in adjacent snapshots are identified as descendants or progenitors if they share at least 50\% of their dark matter particles, and the galaxies they contain are similarly linked.

We identified galaxies in these catalogues as either ``central,'' ``satellite,'' or ``field'' galaxies. Central galaxies are galaxies which share a host halo with at least one other galaxy and are labeled by \citet{DeBla07} as being at the centre of the group's dark matter halo. Satellite galaxies are galaxies which share a host halo with at least one other galaxy and aren't labeled as central galaxies. Field galaxies are galaxies which do not share a host halo with any other galaxy: they are centrals in a halo with no satellites.

\subsection{Models}
\label{sec:Models}

\subsubsection{Halo Models}
\label{sec:halomodels}

For ease of calculations, we assume that all dark matter lies within spherically-symmetric halos\footnote{Triaxial models \citep{JinSut02} provide a more accurate representation of real dark matter halos, but this is not necessary for our purposes. Per \eqref{mainlens}, with a sufficient number of lenses and sources stacked together, the lensing signal of a triaxial halo (which appears elliptical in projection) will be indistinguishable from the lensing signal of a spherically-symmetric halo. This is due to the fact that the noise due to neglecting triaxiality is of order $\sim20\%$ per pair, while shape noise is 1--2 orders of magnitude larger.}. We modeled the mass distributions of all lens halos as truncated NFW profiles, as defined by \citet{BalMarOgu09}, with a density profile:
\begin{equation}
\rho(x) = \frac{M\sbr{0}}{4\pi r\sbr{s}^3}\frac{1}{x(1+x)^2}\frac{\tau^2}{\tau^2 + x^2}\textrm{,}
\label{eq:truncNFW}
\end{equation}
where $ M\sbr{0} = M\sbr{200} \times (\ln(1+c)-c/(1+c))^{-1} $, $ r\sbr{s} = r\sbr{200}/c $, and $ x = r/r\sbr{s} $. $c$ is the concentration parameter, in principle unique to each halo, and $\tau = r\sbr{tidal}/r\sbr{s}$ determines the truncation radius. This model provides an analytic form for the weak lensing signal, as outlined in \citet{BalMarOgu09}. In this paper, we assume that for halos that haven't been tidally disrupted, $\tau=2c$, which implies $r\sbr{tidal}=2r\sbr{200}$. This assumption is consistent with results from \citet{HilWhi10} and \citet{OguHam11}, the latter of which supports a value of $\tau$ between $2c$ and $3c$. We determine $c$ using the mass-concentration relation determined by \citet{NetGaoBet07} for halos in the Millennium Simulation:
\begin{equation}
\label{eq:cfm}
c = 4.67\times\left(\frac{M\sbr{200}}{10^{14} h^{-1} \textrm{M}_{\odot}}\right)^{-0.11}\textrm{.}
\end{equation}

To see how weak lensing signals depend on the mass distribution within groups, we experimented with different methods of apportioning a group's mass to its constituent galaxies. \citet{DeBla07}'s catalogue provides the total mass of a group's halo (which includes the mass of all subhalos) at any snapshot and the stellar masses of all galaxies. Note that the stellar mass is not expected to evolve significantly after a galaxy joins a group \citep{PadLagCor09}, so if there is a monotonic relation between infalling halo mass and stellar mass, the latter can be used to estimate the mass of the galaxy's dark matter halo prior to joining the group.   Here we make this identification using the method of ``abundance matching'', originally applied to entire halos \citep{MarHud02, YanMo03} and later subhalos \citep{ValOst04, ConWecKra06}. We use the following formula, from \citet{GuoWhiLi10}:
\begin{eqnarray}
\label{eq:MhfromMs}
0.129\times\frac{m\sbr{halo}}{m\sbr{stellar}} = \Biggl(\left(\frac{m\sbr{halo}}{10^{11.4} \textrm{M}_{\odot}}\right)^{-0.926}\nonumber \\
+\left(\frac{m\sbr{halo}}{10^{11.4} \textrm{M}_{\odot}}\right)^{0.261}\Biggr)^{2.44}\textrm{.}
\end{eqnarray}
See \appref{MhfMs} for details on how we used this formula to calculate $M\sbr{halo}$ given $M\sbr{stellar}$.\footnote{There is a log-normal scatter in this relation of $\sim0.17$ dex \citep{YanMo12}, which we do not simulate. Testing showed that the effect of this scatter on our results was negligible.} From here on, we will be referring to the halo mass calculated this way as the ``infall mass.'' The mass of any given group is then the total mass of all of its constituent galaxies' halos:
\begin{equation}
\label{eq:MtotfromMhalo}
M\sbr{tot} = \Sigma m\sbr{halo}\textrm{.}
\end{equation}
This prescription, which is based on estimates of halo mass from stellar mass, tends to overestimate the total group mass, as compared with the group masses in \citet{DeBla07}'s catalogue, as \citet{GuoWhiLi10}'s formula was determined based on observational data rather than data from the Millennium Simulation.

\subsubsection{Stripping Models}
\label{sec:strmodels}

Using the data for total group mass and satellite infall mass, we constructed the following three models for the mass distributions of satellite and group halos:

\begin{itemize}
\item Pure Group: A test case in which all group mass is assigned to the central halo, and the masses of all satellite halos are set to zero.
\item No Stripping: A model expected to correspond to unrelaxed groups. Satellites retain their infall masses. The central halo's mass is then set equal to $M\sbr{tot}$ minus the mass in satellite halos.
\item Stripping: A model expected to correspond to relaxed groups, on average. Satellite halos' masses are decreased relative to their infall mass by an amount dependent on their projected distance from the group centre. The central halo's mass is then set equal to $M\sbr{tot}$ minus the mass that remains in satellite halos.
\end{itemize}

We also use the following two datasets for comparison purposes:

\begin{itemize}
\item Matched Field: A selection of field galaxies, matched in redshift and stellar mass to the satellite galaxies used in the above three models. This dataset allows us to measure what the lensing signal around the galaxies in the No Stripping model would look like if the contribution from being in a group environment were removed.
\item Particle Ray-Tracing (PRT): This dataset uses the ray-tracing simulation performed by \citet{HilHarWhi09}, which uses the positions of dark matter particles rather than our halo models. Groups are selected using the group masses in \citet{DeBla07}'s catalogue, rather than our calculated $M\sbr{tot}$.
\end{itemize}

In order to model stripping within groups, we applied results from Fig 15 of \citet{GaoWhiJen04} to estimate a satellite's retained mass fraction from the projected distance between it and its group's centre, using the following function, fit by hand to the figure:
\begin{equation}
\frac{M\sbr{ret}}{M\sbr{init}} \approx -0.464\left(\frac{r}{r\sbr{200}}\right)^2+1.03\left(\frac{r}{r\sbr{200}}\right)+0.058\textrm{.}
\label{eq:MretfromR}
\end{equation}

We modeled the reduction of satellite mass due to stripping by decreasing their tidal radii. The relationship between mass within the virial radius and tidal radius is given by:
\begin{eqnarray}
M\sbr{halo} = M\sbr{0}\frac{\tau^{2}}{\left(\tau^{2}+1\right)^{2}}\left[\left(\tau^{2}-1\right)\ln(\tau)+\tau\pi-\left(\tau^{2}+1\right)\right]\textrm{.}
\label{eq:Mtotfromtau}
\end{eqnarray}
As this equation is not analytically invertible, we use a solution-space search to find the value of $\tau$ which gives the proper value for $\frac{M\sbr{ret}}{M\sbr{init}}$.

As has been shown by \citet{GaoWhiJen04}, the fraction of group mass contained within satellites at a given radius increases with distance from the group centre. This implies that in the Stripping and No Stripping models, we should use a halo density profile for the central halo which converges to the original profile near the core, but is progressively lower than the original profile as the radius increases. We accomplish this with the following, modified version of the truncated NFW profile, which we call the ``contracted'' NFW profile:
\begin{equation}
\rho(x) = \frac{M'\sbr{0} f_r}{4\pi r\sbr{s}'^{3}}\frac{1}{x'(1+x')^{2}}\frac{\tau'^{2}}{\tau'^{2} + x'^{2}}\textrm{,}
\end{equation}
where $ f_r $ is the fraction of mass retained by the group halo, and $ M'\sbr{0} $, $ r'\sbr{s} $, $ x' $, and $ \tau' $ are calculated with new concentration $ c' > c $, which satisfies the equation:
\begin{equation}
\frac{c^2}{\ln(1+c)-c/(1+c)} = \frac{f_r c'^2}{\ln(1+c')-c'/(1+c')}\textrm{.}
\label{eq:cpfromcf}
\end{equation}

\begin{figure}
\centering
\includegraphics[scale=0.4]{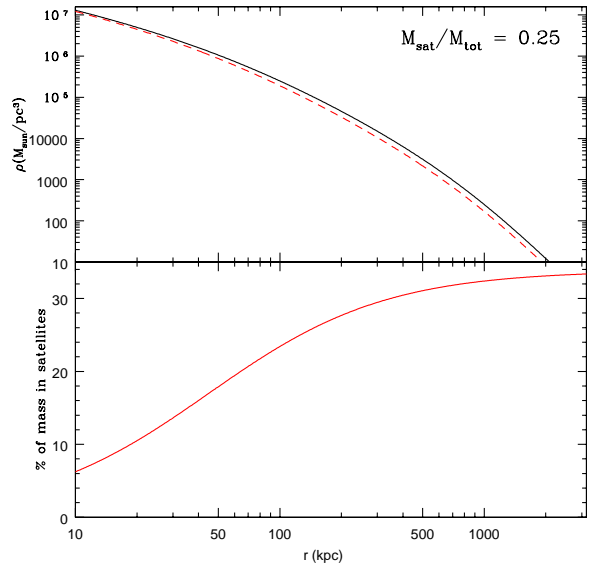}
\caption[]{Top panel: An illustration of our ``contracted NFW'' profile (red dashed) compared to the initial profile (black solid), with a group of $M\sbr{tot} = 10^{13}$~M$_{\odot}$ and $M\sbr{sat}/M\sbr{tot} = 0.25$. The profile converges with the initial NFW profile near the core, but falls below it away from the core. Bottom panel: The resultant fraction of mass assumed to be in satellites at a distance $r$ from the center of this profile. This curve meets our requirements of decreasing to zero at the core and rising with radius.}
\label{fig:NFWmod}  
\end{figure}

This ensures that in the region near the core, where $ x \ll 1 $, this profile will converge with the profile given by \eqref{truncNFW}. Since this profile has a higher concentration than the original profile, the density will decrease faster with radius than in the original profile. There is no strong physical evidence for this specific model; it is used simply because it meets the basic requirements of converging to the initial NFW profile near the core while having a greater fraction of mass in satellites far from the core, and because its lensing signal can be analytically calculated in the same manner as an NFW profile's lensing signal.

An illustration of this transformation, along with a plot of the mass fraction assumed to be contained within satellites under this model, can be seen in \figref{NFWmod}. The weak lensing signal for this model can be calculated similarly to the original model, simply by using a different concentration and applying $ f_r $ as a scaling factor. For consistency, we use this profile for all halos, with $ f_r = 1 $ for all satellite and field halos.

\section{Lensing Simulations}

\label{sec:simulations}
\label{sec:LensingSimulations}

In this section we discuss how we simulate shear from weak gravitational lensing and our methods to adjust the lensing signals around satellite galaxies to remove the contributions from group halos. In \secref{lensoverview}, we provide a brief overview of weak gravitational lensing and some useful formulae used for it. In \secref{Ellipticity}, we explain our algorithm to determine the ellipticities of source galaxies. In \secref{Stacking}, we explain how we analyze the data to calculate lensing signals, and we discuss both of our methods for correcting satellites' lensing signals for the contribution from group halos.

\subsection{Lensing Overview}

\label{sec:lensoverview}

Here we briefly review weak lensing. A more thorough explanation of the tools of weak lensing, with an emphasis on lensing in clusters,  can be found in \citet{KneNat11}, but a brief explanation of the physics and terminology follows: the presence of mass distorts spacetime, causing light to be deflected when it passes near a massive object. In the so-called ``weak lensing'' regime, this manifests as a coherent distortion of the shapes of background galaxies, stretching them tangential to the foreground object, in an effect known as ``shear.'' Although the variation in the intrinsic shapes of background galaxies is significantly larger than the shear caused by foreground objects, the shape of any given background galaxy will still provide an unbiased, albeit low signal-to-noise, estimate of the local shear. This makes it possible to gather usable data by stacking a sufficient number of lens galaxies together.

In practice, shear is usually observed and measured through the dimensionless parameter $\gamma_t$ (known as the tangential shear), but as the magnitude of this parameter depends on the redshifts of both the source galaxy and the foreground object, for this paper we will use the measurement $\Delta\Sigma$, which is related to $\gamma_t$ through:
\begin{equation}
\left<\gamma_t\right> = \frac{\Delta\Sigma}{\Sigma\sbr{crit}}\textrm{,}
\end{equation}
in a circle of any size around any given point, where:
\begin{equation}
\Sigma\sbr{crit} = \frac{c^2 D\sbr{s}}{4\pi G D\sbr{ls}D\sbr{l}}\textrm{.}
\end{equation}
and $D\sbr{s}$ is the angular diameter distance to the source, $D\sbr{l}$ is the angular diameter distance to the lens, and $D\sbr{ls}$ is the angular diameter distance from the lens to the source.

For an arbitrary mass distribution, $\left<\Delta\Sigma\right>$ at a radius $r$ from a central point can be calculated through:
\begin{equation}
\left<\Delta\Sigma(r)\right> = \overline{\Sigma(<r)}-\overline{\Sigma(r)}\textrm{,}
\label{eq:mainlens}
\end{equation}
where $\overline{\Sigma(<r})$ is the surface density averaged for all points contained within this circle, and $\overline{\Sigma(r)}$ is the surface density averaged for all points on the edge of this circle. This prescription works even for mass distributions that are not axisymmetric, as long as all points in a given annuli around a lens object are stacked together.

\subsection{Determining Ellipticity}
\label{sec:Ellipticity}

While full ray-tracing, such as that performed by \citet{HilHarWhi09}, is a powerful tool for estimating weak lensing signals caused by an arbitrary mass distribution, it has a few limitations. The accuracy of the ray-tracing is limited by the resolution of the simulations used to generate the lens mass distribution, which causes the algorithm to noticeably underestimate the lensing signal within ~50 kpc of halo cores (this effect can be seen in \figref{sc_cen}). Additionally, the algorithm has no method to easily model different hypothetical mass distributions, as we wish to test here. As such, we've developed a modified ray-tracing algorithm, in which we assume that all dark matter lies in spherical halos in amounts and distributions dependant on the models described in \secref{strmodels}. The algorithm proceeds as follows:

\begin{enumerate}
\item Lens galaxies are assigned mass, concentration, tidal radius, and scaling factor under one of the models described in \secref{strmodels}.
\item In order to eliminate edge effects, the catalogue of lens galaxies used for this algorithm includes galaxies within a wider field of view than the catalogue of source galaxies. Since the fields are $4 \times 4$ degrees, we accomplish this by only using source galaxies within a central 1.4 x 1.4 degree square field, which also significantly reduces the amount of computational time needed for later analysis. For the rest of our analysis, we also only use lens galaxies that lie within this central field.
\item Each source galaxy is initialized with zero ellipticity in both components (shape noise is simulated at a later stage of analysis).
\item For each source galaxy, tangential shear is applied to it for every lens galaxy where $ \left(\left[\textrm{RA}\sbr{lens} - \textrm{RA}\sbr{source}\right] \times\cos(\textrm{Dec}\sbr{lens})\right)^2 + \left(\textrm{Dec}\sbr{lens}-\textrm{Dec}\sbr{source}\right)^2 < \left(\Delta\sbr{max}\right)^2 $ and $ z\sbr{lens} < z\sbr{source} $. $ \Delta\sbr{max} $ is an upper limit to conserve computational time and ensure that the rectangular boundaries of the field will not cause artifacts in the lensing signal. The strength of the shear applied is determined using \citet{BalMarOgu09}'s calculations for a truncated NFW halo and our modifications for a contracted NFW profile (see \secref{strmodels}).
\item Shear is added linearly to the source galaxy's ellipticity for all lens galaxies within this lightcone.
\end{enumerate}

This process in principle accounts for the two-halo term seen in the lensing signal around groups, which is caused by other nearby groups. However, our application of a cut-off angle $\Delta\sbr{max}$ suppresses lensing signals at large radii, meaning the two-halo term is not observable in our simulated lensing signals. This method does not account for dark matter that is correlated with the positions of galaxies and groups but not considered part of their halos, which also results in a slight suppression of the lensing signal at large ($R\gtrsim 1000$~kpc) scales. Since we are interested in the stripping of subhalos, for which the relative signal is strongest at intermediate ($50$~kpc $\lesssim R \lesssim 400$~kpc) scales, this suppression at large scales will have no effect on our analysis.

\subsection{Stacking Lensing Signals}
\label{sec:Stacking}

Once the source catalogue has been prepared with simulated ellipticities for each source, we proceed to stack the lensing signal around a sample of lenses in order to observe the average signal for the sample. We separate our sample of lens galaxies into samples of central galaxies and samples of satellite galaxies.

With satellite galaxies, we are interested in comparing the lensing signals of similar satellites in groups of different masses. For the sake of clarity in discussing these limits, we use $m\sbr{min} $ and $m\sbr{max} $ to define the lower and upper limits, respectively, for satellite mass in our samples, and $M\sbr{min} $ and $M\sbr{max} $ to define the lower and upper limits, respectively, for the masses of the groups in which they reside.

With each subset of lens galaxies, we calculate the average tangential shear within radial bins. The results of these calculations and analysis of the results can be seen in \secref{SigCen} and \secref{SigSat}. The subsets of satellites will typically show some contribution from their groups' halos to the lensing signals around them. Since the observed lensing signal around satellites is the combination of the contributions from the satellites' halos and their groups' halos:
\begin{equation}
\Delta\Sigma\sbr{obs} = \Delta\Sigma\sbr{group}+\Delta\Sigma\sbr{sat}\textrm{,}
\end{equation}
we can isolate the portion of the signal contributed by the satellites' halos ($\Delta\Sigma\sbr{sat}$) by estimating the groups' contribution to the lensing signal ($\Delta\Sigma\sbr{group}$) and subtracting it from the observed lensing signal ($\Delta\Sigma\sbr{obs}$). We have applied the two methods detailed below to do this, and we compare their results in \secref{SigIdeal} and \secref{rdat}. 

\subsubsection{Mirror Method}
\label{sec:Mirror}

\begin{figure}
\centering
\includegraphics[scale=0.4]{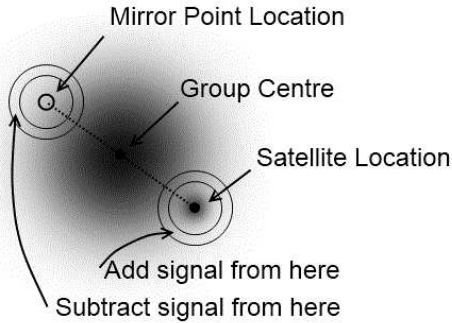}
\caption[Ilustration of the Mirror Method]{An illustration of the Mirror Method. For each satellite used as a lens, we also calculate the lensing signal around a point opposite the group centre from this satellite. We subtract this signal from the signal around the satellite to correct for the contribution of the group's halo to the lensing signal. See also Figure 1 of \citet{PasHilHar11} and related discussion.}
\label{fig:mir_ilus}  
\end{figure}

The first method for isolating the satellite's contribution to the lensing signal involves using a ``mirror'' point in the group to measure the group's contribution to the satellite's lensing signal. This sample point is placed opposite the group centre from the satellite, at the same distance from it, as illustrated in \figref{mir_ilus}. In an ideal scenario, the lensing signal around this sample point will include exactly the same contribution from the group's halo as at the satellite's location, while including little contribution from the satellite itself.

This method is in practice identical to the ``Galaxy-Galaxy Lensing with Calibration'' method used by \citet{PasHilHar11}. Figures 1 and 2 in that paper provide a useful illustration of how this method functions, and discussion of it is found in Section 3.

\subsubsection{Ensemble Method}
\label{sec:Signal}
\label{sec:Ensemble}

The second method involves estimating the masses of groups and their contributions to the lensing signals around their satellites. First, we create a sample of groups with $ M\sbr{min} < M < M\sbr{max} $ and each with at least one satellite with $ m\sbr{min} < m < m\sbr{max} $. We use a weighted average of the groups' masses to generate an ``ensemble'' halo, representative of the sample. We then assign this ensemble halo a concentration based on \eqref{cfm}. For each source in the sample, we then use our ensemble halos and the positions group centres to calculate the group's expected contribution to the source's ellipticity and subtract off this amount.

\section{Lensing Analysis}
\label{sec:analysis}

In this section we present the lensing signals that result from our simulations and discuss the effects of various sources of error on these signals. In \secref{errorsources}, we discuss the dominant sources of error in this type of analysis. In \secref{SigCen}, we present the lensing signals that result when group centres are stacked together. In \secref{SigSat}, we present the lensing signals that result when satellite galaxies are stacked together, we show the effects of our methods to reduce the contributions of group centres to these signals, and we investigate the effects of centering and group mass errors on these signals.

\subsection{Dominant Sources of Error}
\label{sec:errorsources}

In order to properly assess whether our methods for measuring satellites' lensing signals are viable with observational data, it is necessary to account for the sources of error present in such datasets. In this section, we explain the major sources of error we expect in datasets and how we replicate them in our simulations.

\subsubsection{Shape Noise}
\label{sec:survey}

There is significant variation in the orientation of galaxies in the sky, independent of the effects of gravitational lensing. The distribution of the unlensed shapes of background galaxies can be approximately modeled by a Gaussian distribution for both components of ellipticity, with a mean of zero and a standard deviation of $0.28$. This value was calculated from the distribution of ellipticities measured in background galaxies in the CFHTLenS, estimated with the lensfit method \citep{MilKitHey07,KitMilHey08,HeyMil12}.

We account for the impact of shape noise on our measurements by setting the size of the error bars in our plots to account for its expected impact. This allows us to precisely predict the mean expected measurements, unaffected by shape noise, and to estimate the scatter that will be seen in observational data. Our error estimates assume data from a survey similar to the overlap of the CFHTLenS survey and the GAMA-II survey. This hypothetical survey covers 50 deg.$^2$ of the sky, with an effective $15$ sources per arcmin.$^2$ and galaxy spectra complete to $r < 19.8$.

\subsubsection{Group Mass and Group Assignment Errors}

With observational data, group masses must be estimated through one of various methods, such as using the velocity dispersion of constituent galaxies, the luminosity of the group, X-ray emissions, or lensing signals. In the GAMA-II survey, group masses will be estimated primarily through velocity dispersions, so this is the method that we simulate here. To do this, we first determine which of a group's galaxies are within the survey's detection limits. For each of these galaxies, we generate a random line-of-sight velocity from a Gaussian distribution, where the dispersion is determined by the group's total mass, as given in \citet{DeBla07}'s catalogue. For all calculations relating $\sigma\sbr{1}$ and $M\sbr{200}$, we assume a Singular Isothermal Sphere (SIS) profile, which gives us the relation:
\begin{equation}
\sigma\sbr{1} = \left(10GhM\sbr{200}\right)^{1/3}/\sqrt{3}\textrm{.}
\end{equation}

When groups are detected through a friends-of-friends algorithm, it is expected that some portion of the galaxies believed to be within a group will in fact be ``interlopers.'' This happens when field galaxies lie at a similar position in the sky as a group at a different cosmological redshift, but with a peculiar velocity that makes their apparent redshift lie within the redshift distribution of galaxies within the group. For a given FoF-detected group, typically $20\%$ of the galaxies identified as belonging to it are in fact interlopers \citep{RobNorDri11}.

To simulate the effects of interlopers on group mass estimates, we first determine the number of interlopers that a given group contains. Since the number of interlopers should be proportional to the projected area of the group, we set the mean number of interlopers in a given group proportional to $N\sbr{gal}^{2/3}$. We then use a Poisson distribution to generate the number of interlopers for each group, normalized so that a group with $10$ members will have an expected $2.5$ interlopers (resulting in a $20\%$ interloper fraction). For each of these interlopers, we then generate a velocity for them, drawn from a uniform distribution, $v\sbr{min}-v\sbr{tol} < v\sbr{rand} < v\sbr{max}+v\sbr{tol}$, where $v\sbr{min}$ and $v\sbr{max}$ are the minimum and maximum velocities of real members of this group, and $v\sbr{tol}$ is the velocity tolerance used by the FoF algorithm. In the algorithm used for the GAMA-II survey, $v\sbr{tol}$ is not fixed, and in fact depends on the luminosities of the galaxies being linked, but for simplicity's sake, we use a constant value of $150$ km/s here.

Once velocities for real members and interlopers are generated, we then calculate the predicted group mass using these velocities. When this mass is used in our analysis, we use the subscript ``dyn.''

The presence of interlopers will also affect the lensing signal for the ``Stripping'' model, as interlopers would typically be field galaxies, which will exhibit an unstripped mass distribution. To simulate this effect, we recalculate the signal in the ``Stripping'' model as:
\begin{equation}
\Delta\Sigma\sbr{S,new} = 0.8\times\Delta\Sigma\sbr{S,old} + 0.2\times\Delta\Sigma\sbr{NS}\textrm{,}
\end{equation}
where $\Delta\Sigma\sbr{S}$ is the lensing signal found with the ``Stripping'' model, and $\Delta\Sigma\sbr{NS}$ is the lensing signal found with the ``No Stripping'' model.

We use both the modifications to group mass, which affects the binning of centres, and the simulation of the contamination caused by interlopers only for the plots shown in \secref{rdat}, which attempt to predict the strength of the signal with a CFHTLenS + GAMA-II-like survey (see \secref{survey}).

\subsubsection{Centering Errors}
\label{sec:Centering}

It is not trivial to determine the position of a group's centre. Since the majority of the mass in a group is dark matter, and galaxies are an imperfect tracer of the dark matter distribution, any method which uses galaxies' positions to estimate the centre of a group's dark matter halo will have some degree of error. The amount of error can be roughly estimated by analysis of the velocity dispersion about suspected centres, as done by \citet{SkiYan11}; by comparing the results of multiple independant methods for identifying the group's centre, or by fitting the lensing signal around a stack of groups with a profile convolved with error, as done by \citet{GeoLeaBun12}; or through direct analysis of N-body simulations, as done by \citet{JohSheWec07}. While \citet{JohSheWec07}'s analysis of simulations implies that rich groups are less likely to be miscentered than poor groups, this doesn't seem to be the case with real groups. Recent work by \citet{SkiYan11} and \citet{GeoLeaBun12} all show that this trend is in fact reversed in real groups. For the most massive groups, the centre is misidentified roughly $40\%$ of the time, implying that misidentifying the BCG is a significant problem. \citet{GeoLeaBun12}'s work has also demonstrated that using either the most massive or brightest galaxy in a group as the centre is the optimal method when X-ray observations aren't available.

Therefore, to simulate the effects of centering errors, we assume that all groups are miscentered according to a 2D Gaussian with $\sigma$ proportional to their scale radii $r\sbr{s}$. When groups are stacked together, this leads to roughly the same net effect as observed previously, but it allows us to adjust the magnitude of the effect by altering only the constant of proportionality (rather than both the fraction of groups in which the centre is misidentified and the dispersion of the false centres relative to the real centres), to easily compare different amounts of error. This does not, however, perfectly match the best fit models of \citet{GeoLeaBun12}, so only qualitative conclusions should be made from our simulations involving centering error. We compare models with centering errors $\sigma = 0$, $r\sbr{s}$, and $2r\sbr{s}$, which covers the expected range of error. When real data are available, it will be necessary to determine the proper amount of error through an analysis of the lensing signals around suspected group centres, and then recalibrate our predictions for it, using a method similar to \citet{GeoLeaBun12}'s method of fitting the lensing signals with an NFW profile convolved with centering error.

For the Mirror Method, centering errors will have an impact because if there's error in the position of the group's centre, there will also be error in the position of the mirror point. On average, this error will cause the mirror point to lie farther away from the group centre than it actually should, and so it underestimates the group's contribution to the lensing signal around its satellite.

For the Ensemble Method, centering errors will affect the calculated signal from the ensemble halo. The reason for this is that errors in the positions of group centres will suppress the group signal at low radii ($R\lesssim\sigma\sbr{cen}$), thus causing a lower mass to be fit. They will also change the calculated satellite distances from group centres, increasing them more often than decreasing them, and thus decreasing the ensemble halo's typical impact on source ellipticity. In practice, this effect is somewhat smaller than the effect of centering errors on the Mirror Method (see \secref{rdat} for details).

\subsection{Lensing Signal around Group Centres}
\label{sec:SigCen}

\begin{figure}
  \centering
  \includegraphics[scale=0.4]{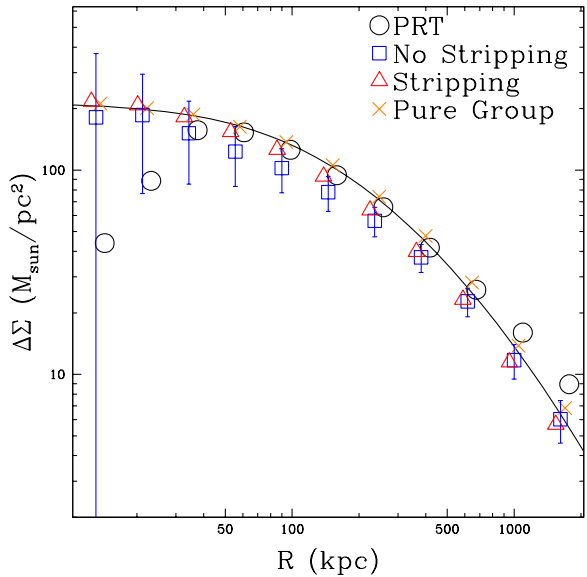}
  \caption[Lensing signal around group centres for all four mass models and the PRT data.]{Lensing signal, measured with $\Delta\Sigma$, around a sample of galaxy groups with $ 10^{13}$~M$_{\odot} < M\sbr{tot} < 10^{14}$~M$_{\odot} $ drawn from the Millennium Simulation, with no centering or group mass errors applied. All four mass models detailed in \secref{Models} are shown here, along with the lensing signal that results from using the particle ray-tracing (PRT) shear data. The solid line shows the predicted shear for a group which has the weighted average mass of the sample ($\overline{M\sbr{tot}} = 4.00 \times 10^{13}$~M$_{\odot}$) at the weighted average redshift of the sample ($\bar{z}\sbr{gr} = 0.18$). Error bars are shown for the No-Stripping model which represent the projected errors for data from a hypothetical CFHTLenS + GAMA-II-like survey (see \secref{survey}).}
  \label{fig:sc_cen}
\end{figure}

Since the expected form of the lensing signal around group-centres is well-understood, measuring this signal in our simulations provides a useful test to confirm that our models behave as expected. \figref{sc_cen} shows the average lensing signals around a sample of group centres with $ 10^{13}$~M$_{\odot} < M\sbr{tot} < 10^{14}$~M$_{\odot} $, where $M\sbr{tot}$ is the total mass of the group, for all of our mass models and the particle ray-tracing data, with no centering or group mass errors applied. Our models show rough agreement with the PRT lensing signal at large radii, although the PRT signal drops significantly below the models' signals for sources within $ \approx50$~kpc. This effect was noticed by \citet{HilHarWhi09}, and is due to force softening used within the Millennium Simulation, which flattens the cusps of halos, as well as the smoothing applied during the ray-tracing to suppress particle shot noise.

All of our models appear similar in this regime, with a slight tendency for the ``No Stripping'' and ``Stripping'' models to fall slightly below the ``Pure Group'' model. This implies that our contracted NFW profile is not a perfect model for the background halo of a group with significant substructure in it. More work will be required to determine a better model for this. This effect is minor, however, and all lensing signals are consistent with an NFW profile. If centering errors (see \secref{Centering}) are present in observational data, the lensing signal around group centres should be suppressed at low radii.

\subsection{Lensing Signal around Satellites}
\label{sec:SigSat}

\begin{figure}
  \centering
  \includegraphics[scale=0.4]{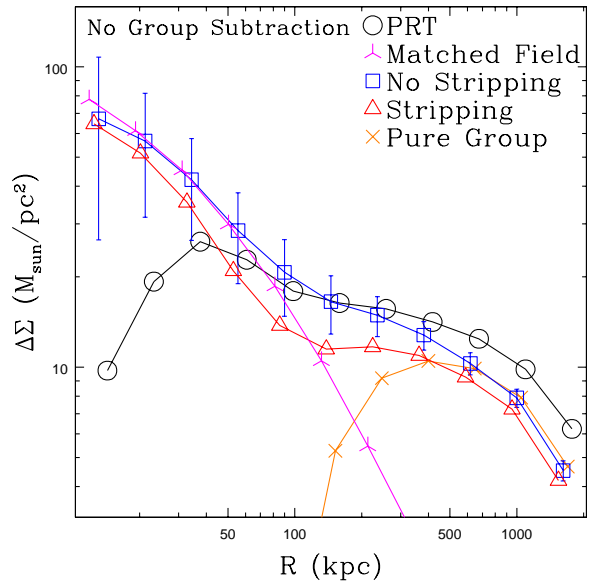}
  \caption[Lensing signal around satellite galaxies for all four mass models and the PRT data.]{Lensing signal, measured with $\Delta\Sigma$, around a sample of satellite galaxies with $ 10^{10}$~M$_{\odot} < m\sbr{halo}$ (as determined with the ``No Stripping'' mass model), and $ 10^{13}$~M$_{\odot} < M\sbr{tot} < 10^{14}$~M$_{\odot}$, drawn from the Millennium Simulation. All three mass models detailed in \secref{Models} are shown here, along with the ``Matched Field'' dataset and the lensing signal that results from using the particle ray-tracing shear data with a similarly-cut sample of lenses. Error bars are shown for the No-Stripping model which represent the projected errors for data from a hypothetical CFHTLenS + GAMA-II-like survey (see \secref{survey}).}
  \label{fig:sc_sat}
\end{figure}

\figref{sc_sat} shows the average lensing signals around a sample of satellite galaxies with $ 10^{10}$~M$_{\odot} < m\sbr{halo}$, where $m\sbr{halo}$ is as used in \eqref{MhfromMs}, and $ 10^{13}$~M$_{\odot} < M\sbr{tot} < 10^{14}$~M$_{\odot}$, for all of our mass models and the particle ray-tracing data. The differences between our models are in principle detectable with the amount of data available in our hypothetical CFHTLenS + GAMA-II-like survey (see \secref{survey}). At radii larger than $ \sim 50$~kpc, the PRT signal lies close to our No Stripping model, which is an encouraging sign that our models are giving reasonable results. However, our ``No Stripping'' model is noticeably different from the ``Matched Field'' dataset, due to the contributions of the satellites' host groups to the lensing signal, and this difference is greater than the difference between it and the ``Stripping'' model. The ``Pure Group'' model, as shown here, shows only the lensing signal that results from setting the satellite masses to zero in our simulations, and so it can be used as an approximation of the group halos' contribution to the lensing signal around satellites. In order to use the ``Matched Field'' dataset as a standard for comparison, it will be necessary to eliminate the groups' contributions to the lensing signals measured here. This may be possible through one of the methods outlined in \secref{Stacking}. Below we have presented the results of our tests of these methods.

\subsubsection{Idealized Group Signal Subtraction}
\label{sec:SigIdeal}

\begin{figure*}
  \centering
  \includegraphics[scale=0.4]{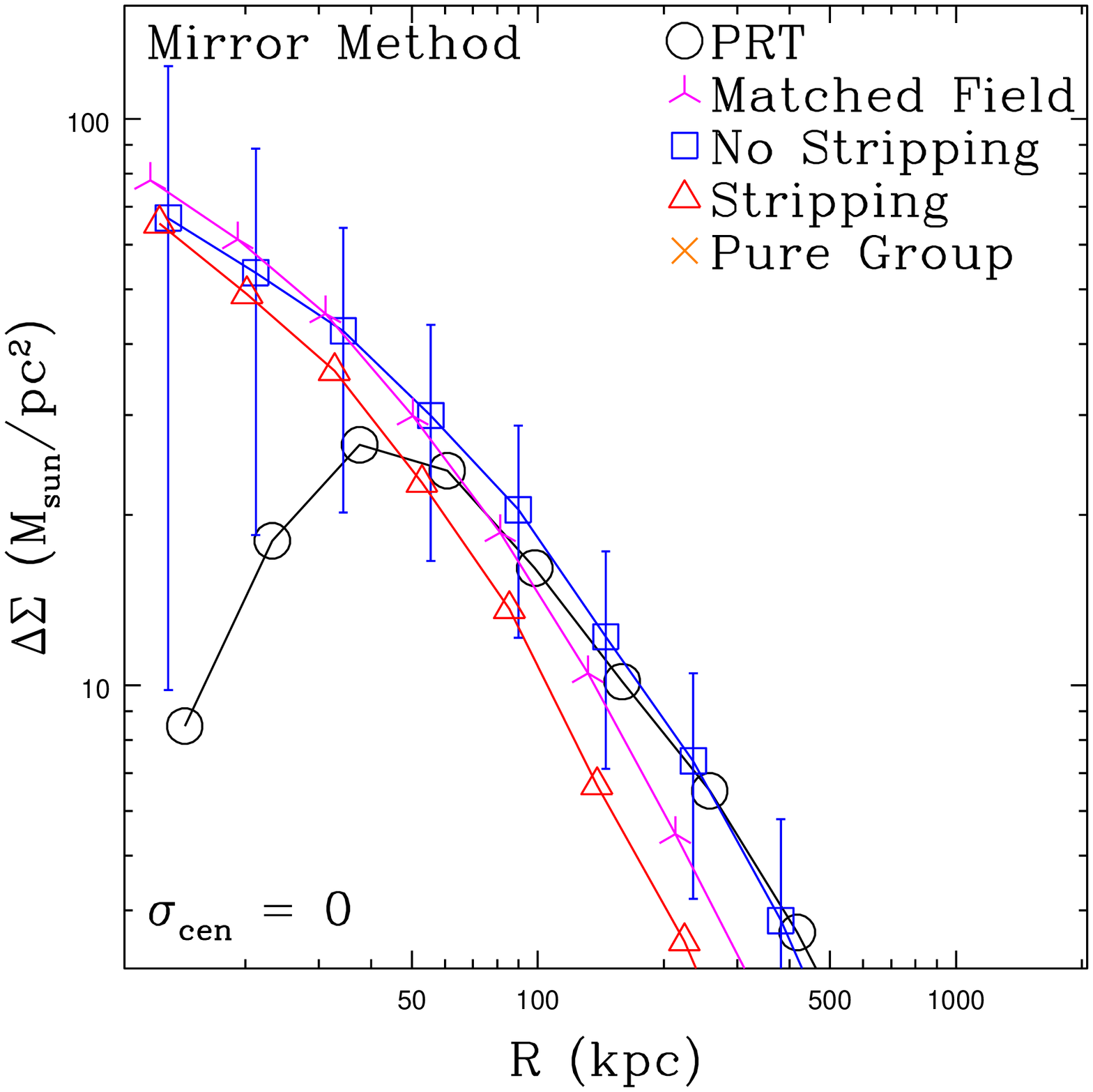}
  \includegraphics[scale=0.4]{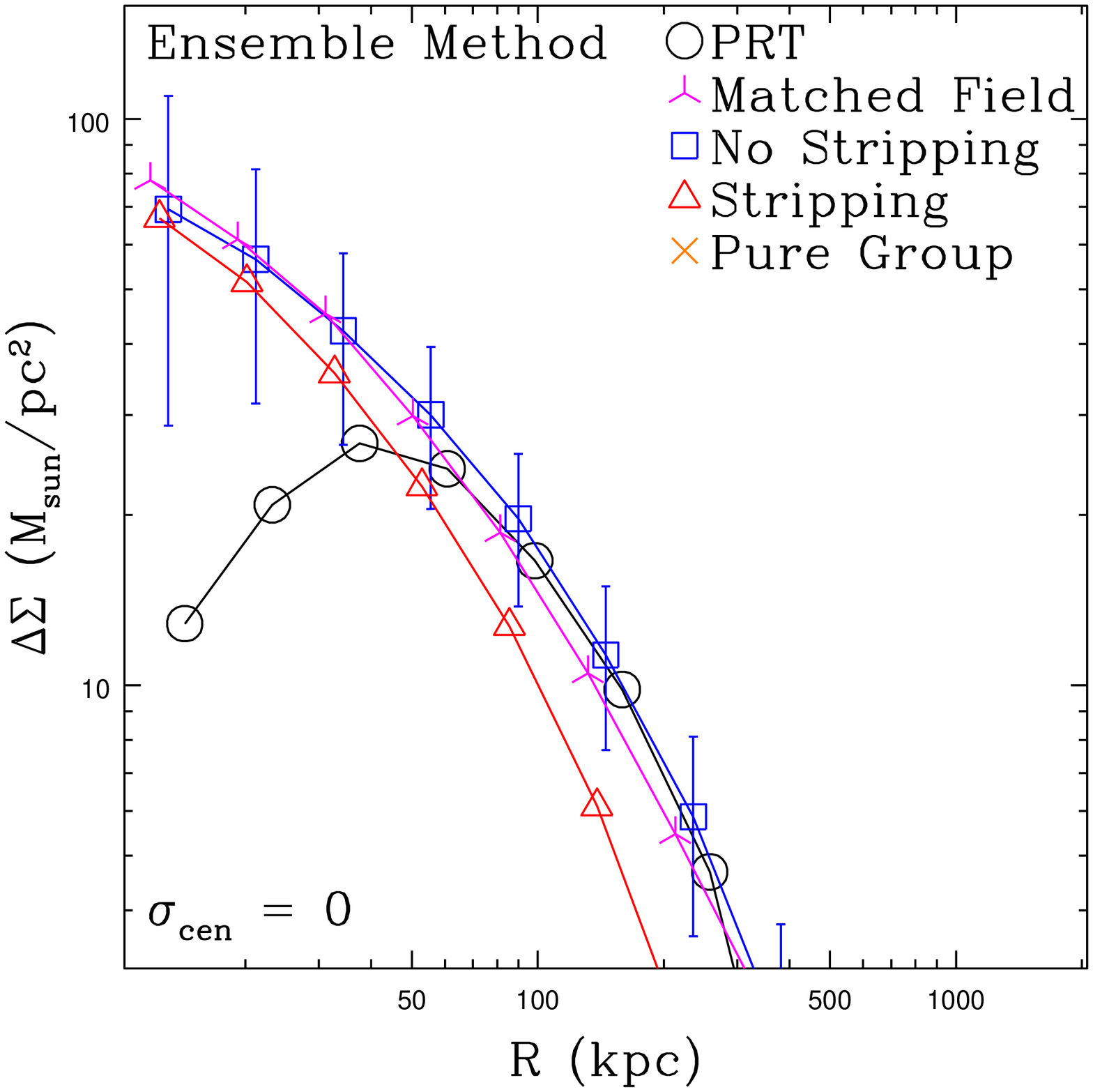}
  \hfill
  \caption[Lensing signal around satellite galaxies for all four mass models and the PRT data, modified through our Mirror Method.]{Same as \figref{sc_sat}, except with the Mirror Method (left panel) and Ensemble Method (right panel) applied to reduce contribution to the lensing signal from groups' halos (except with the ``Matched Field'' dataset).}
  \label{fig:sc_mirror}
  \label{fig:sc_sig}
\end{figure*}

As detailed in \secref{Mirror}, the Mirror Method uses a sample ``mirror'' point opposite the group centre from a satellite to estimate the group's contribution to the satellite's lensing signal. \figref{sc_mirror} shows the result of applying this to a sample of satellite galaxies with $ 10^{10}$~M$_{\odot} < m\sbr{halo}$ and $10^{13}$~M$_{\odot} < M\sbr{tot} < 10^{14}$~M$_{\odot}$, in the idealized scenario in which there are no errors in the positions of group centres or the estimated masses of groups. As can be seen in the figure, the curve for the ``No Stripping'' model is very similar to that of the ``Matched Field'' dataset, and the ``Pure Group'' model has disappeared off the bottom of the plot (it is consistent with zero). This implies that the method does a satisfactory job of eliminating the groups' contribution to the lensing signal around their satellites. The signal-to-noise in the difference between the ``No Stripping'' and ``Stripping'' models is quite strong.

The Ensemble method, explained more thoroughly in \secref{Ensemble}, uses the average mass of the groups in the sample as an average ``Ensemble'' halo. For each group centre, its predicted shear is then subtracted from the ellipticities of all nearby source galaxies. After this, satellites are stacked together, and the lensing signal around them, using the altered ellipticities, is measured. \figref{sc_sig} shows the result of applying this method to a sample of satellite galaxies with $ 10^{10}$~M$_{\odot} < m\sbr{halo}$ and $10^{13}$~M$_{\odot} < M\sbr{tot} < 10^{14}$~M$_{\odot}$, also in the idealized scenario of no centering or group mass errors. The signals from the ``No Stripping'' model show good agreement with the ``Matched Field'' model. Notably, the difference between the ``Stripping'' and ``No Stripping'' models here is greater than when the Mirror Method is used. This is because the Mirror Method results in some of the lensing signal around satellites being lost, particularly for satellites near the centres of groups. Although the Ensemble Method requires more assumptions about the physical properties of the groups, it does not face this problem, and so the resultant signal-to-noise in the difference between the two models is larger.

\subsubsection{Realistic Group Signal Subtraction}
\label{sec:rdat}

\begin{figure*}
  \centering
  \includegraphics[scale=0.4]{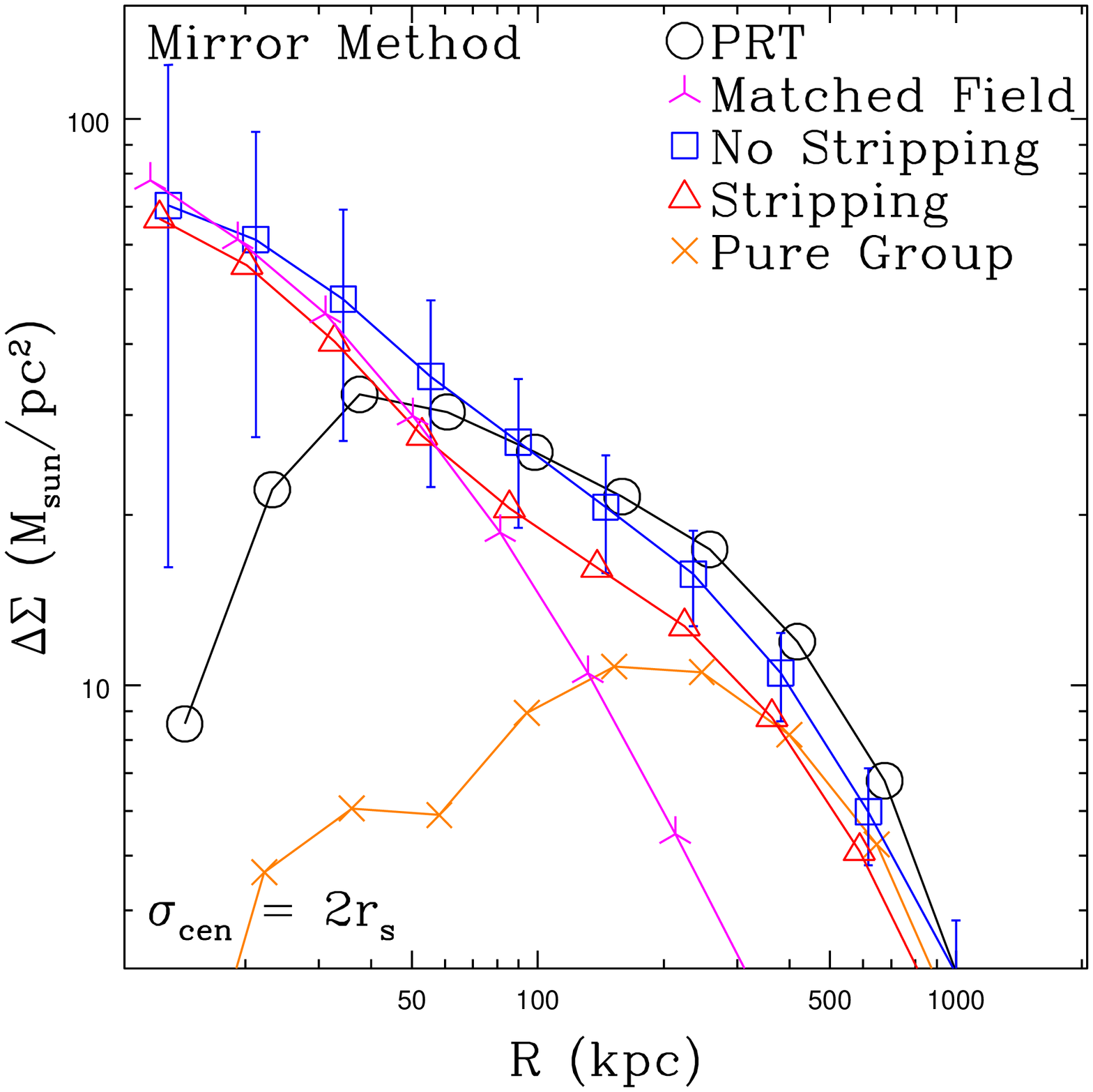}
  \includegraphics[scale=0.4]{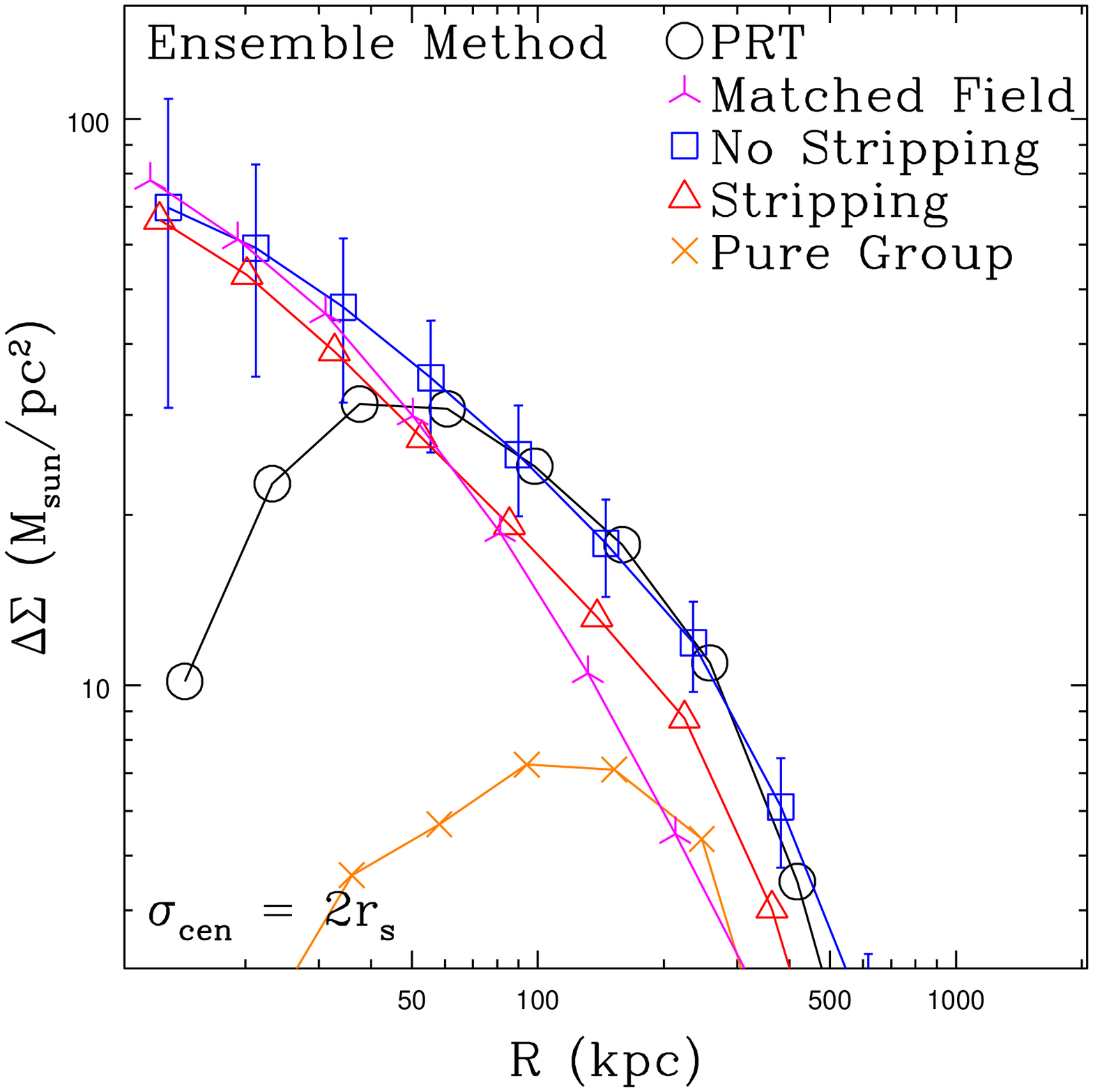}
  \hfill
  \caption[Lensing signals around satellite galaxies for all four mass models and the PRT data, modified through our Mirror Method.]{Same as \figref{sc_sat}, except with the Mirror Method (left panel) and Ensemble Method (right panel) applied to reduce contribution to the lensing signal from groups' halos (except with the ``Matched Field'' dataset), assuming centering errors of $\sigma = 2r\sbr{s}$.}
  \label{fig:sc_mirrore}
  \label{fig:sc_sige}
\end{figure*}

With observational data, the positions of group centres will not be precisely defined, so it is necessary to assess the impact of centering errors on the utility of this method. Figure \ref{fig:sc_mirrore} shows how the lensing signal appears if centering errors of $\sigma\sbr{cen} = 2r\sbr{s}$ are assumed. We simulated this by applying a 2-dimensional Gaussian deviate to the positions of groups centres. This amount of centering error is significantly higher than the errors found by \citet{GeoLeaBun12}, which were typically of order $\sigma\sbr{cen} \approx 0.5 r\sbr{s} - 1 r\sbr{s}$, and so this provides an upper bound to the amount of centering error that might be seen in observational data. As can be seen through comparison with \figref{sc_mirror}, centering errors have a significant impact on the resultant lensing signal.

\begin{figure*}
  \centering
  \includegraphics[scale=0.4]{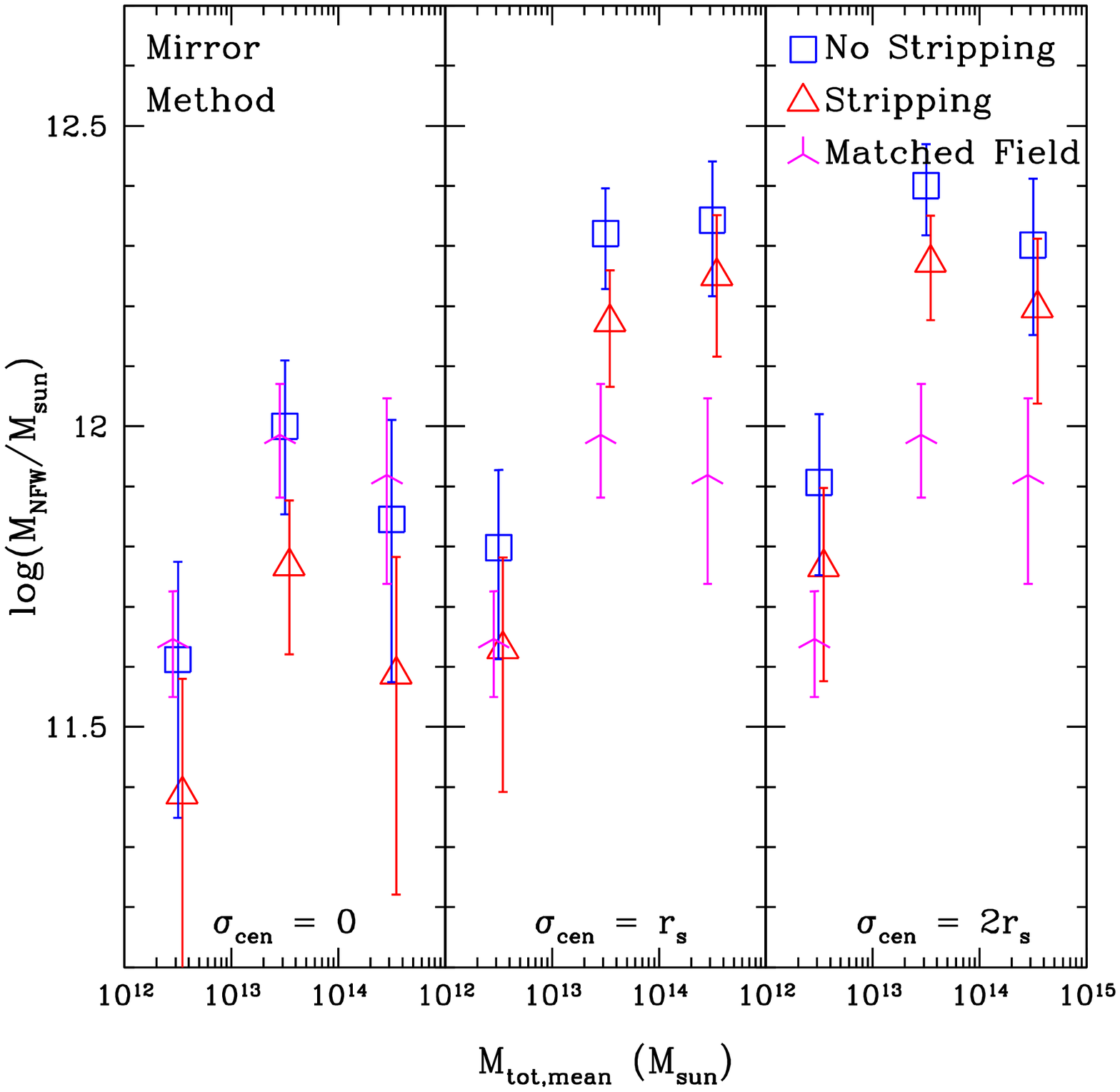}
  \includegraphics[scale=0.4]{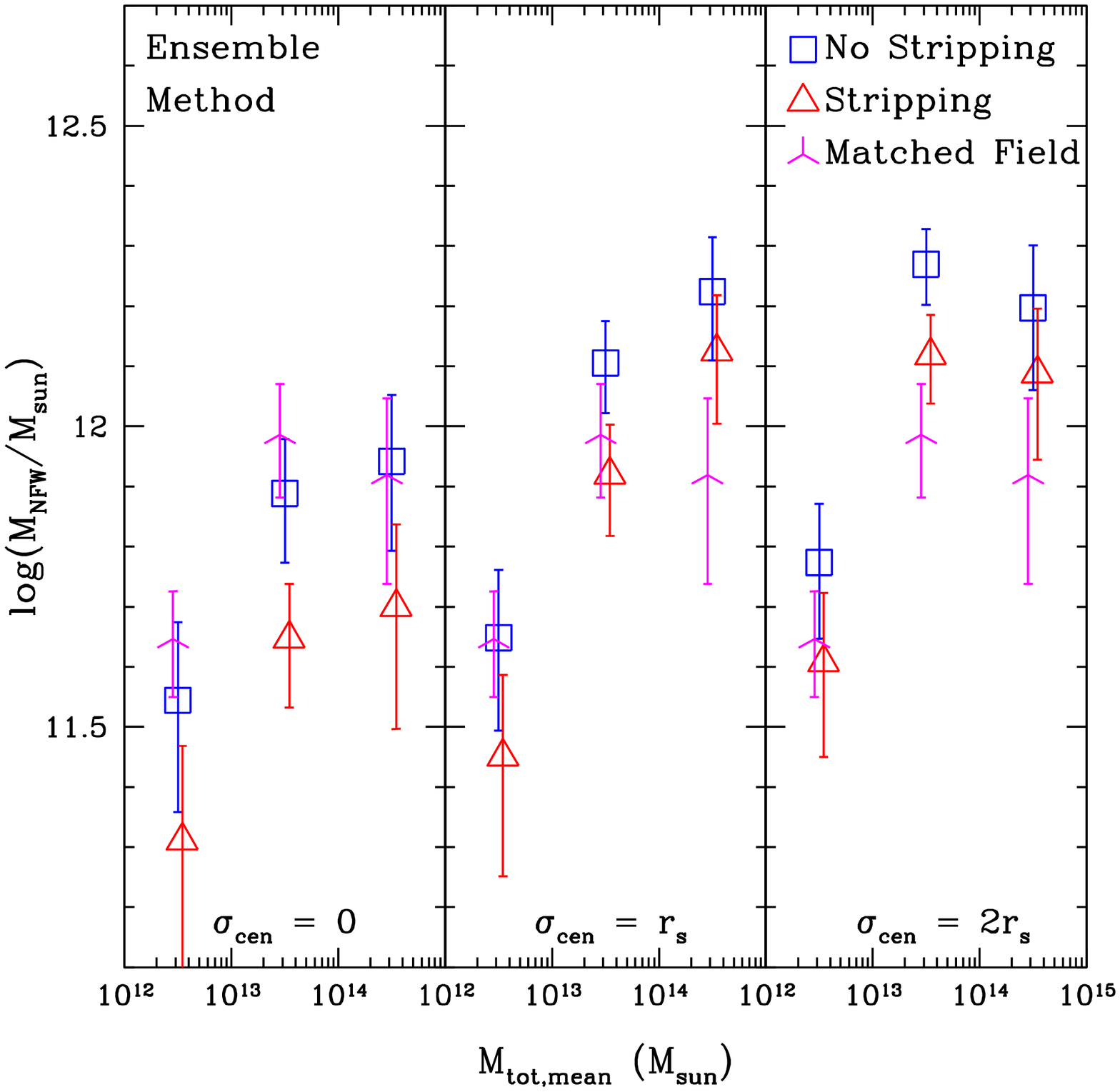}
  \hfill
  \caption[Comparison plot for the best-fit NFW masses with the Mirror Method for various group mass ranges and levels of centering error.]{Best-fit NFW masses for a sample of satellite galaxies with $ 10^{10}$~M$_{\odot} < m\sbr{halo}$ (as determined with the ``No Stripping'' mass model) drawn from the Millennium Simulation, using only source galaxies within 10~to 400~kpc of the lens galaxies. Satellites from three ranges of group mass are included, shown at the average mass for the range, and the Mirror Method (left panel) and Ensemble Method (right panel) have been applied with three different levels of centering error assumed. Centering errors used were $\sigma\sbr{cen} = 0$ (left), $\sigma\sbr{cen} = r\sbr{s}$ (centre), and $\sigma\sbr{cen} = 2r\sbr{s}$ (right). Error bars shown are the projected errors for data from a hypothetical CFHTLenS + GAMA-II-like survey (see \secref{survey}). The ``Matched Field'' mass is observed to increase with group mass as more massive groups typically contain more massive satellites, and these more massive satellites are then matched to more massive field galaxies.}
  \label{fig:sc_mirrorcomp}
  \label{fig:sc_sigcomp}
\end{figure*}

\begin{table*}
\centering
\caption{Statistics for our galaxy samples and NFW profile fits, using the Ensemble Method and assuming $\sigma\sbr{cen}=0$. All masses are in units of $10^{10}$~M$_{\odot}$. S/N$\sbr{F-S}$ is the signal-to-noise ratio in the difference between the fitted masses for the ``Field'' and ``Stripping'' models, and S/N$\sbr{NS-S}$ is the same for the difference between the ``No Stripping'' and ``Stripping'' models.}
\begin{tabular}{|l l l l l l l l l l l l l l l l l|}
 & & \multicolumn{4}{c}{Matched Field} & & \multicolumn{4}{c}{No Stripping} & \multicolumn{2}{c}{Stripping} \\ \hline
$\log{M}$ & & $\overline{z}$ & $\overline{m\sbr{halo}}$ & $m\sbr{fit}$ & $m\sbr{fit,err}$ & & $\overline{z}$ & $\overline{m\sbr{halo}}$ & $m\sbr{fit}$ & $m\sbr{fit,err}$ &  $m\sbr{fit}$ & $m\sbr{fit,err}$ & & S/N$\sbr{F-S}$ & S/N$\sbr{NS-S}$ \\ \hline
12--13 & & .15 & 42 & 45 & \phn9.4 & & .14 & 47 & 34 & 13 & 18 & 11 & & 1.43 & 2.11 \\
13--14 & & .18 & 93 & 97 & 18  & & .16 & 93 & 77 & 19 & 43 & 15 & & 1.88 & 2.33 \\
14--15 & & .14 & 64 & 82 & 25 & & .12 & 70 & 90 & 27 & 47 & 21 & & 1.61 & 1.38 \\
\hline
\end{tabular}
\label{tab:tab1}
\end{table*}

To determine whether or not we will be able to detect a difference between the ``Stripping'' and ``No Stripping'' models when centering errors are present, we've plotted the best-fit NFW mass (determined by minimizing the $\chi^2$ statistic for the group-subtracted lensing signal around satellites in radial bins between 10 and 400 kpc) for our ``Stripping'' and ``No Stripping'' models in both group mass bins in Figure \ref{fig:sc_sigcomp}, for varying amounts of centering error. As this figure shows, centering errors increase the best-fit masses in all mass bins, but it is only a significant problem for the more massive groups. If we have no knowledge of how large centering errors in observational data will be, the uncertainty they cause will be too large for us to discern between stripped and unstripped models. This is unlikely to be the case, however, as we will be able to measure the lensing signals around group centres to estimate the magnitude of centering errors through a fit such as that performed by \citet{GeoLeaBun12}, and significant amounts of error will manifest as a much-poorer $\chi^2$ fit. Details for all models, assuming no centering error, can be seen in \tabref{tab1}.

\begin{figure*}
  \centering
  \includegraphics[scale=0.4]{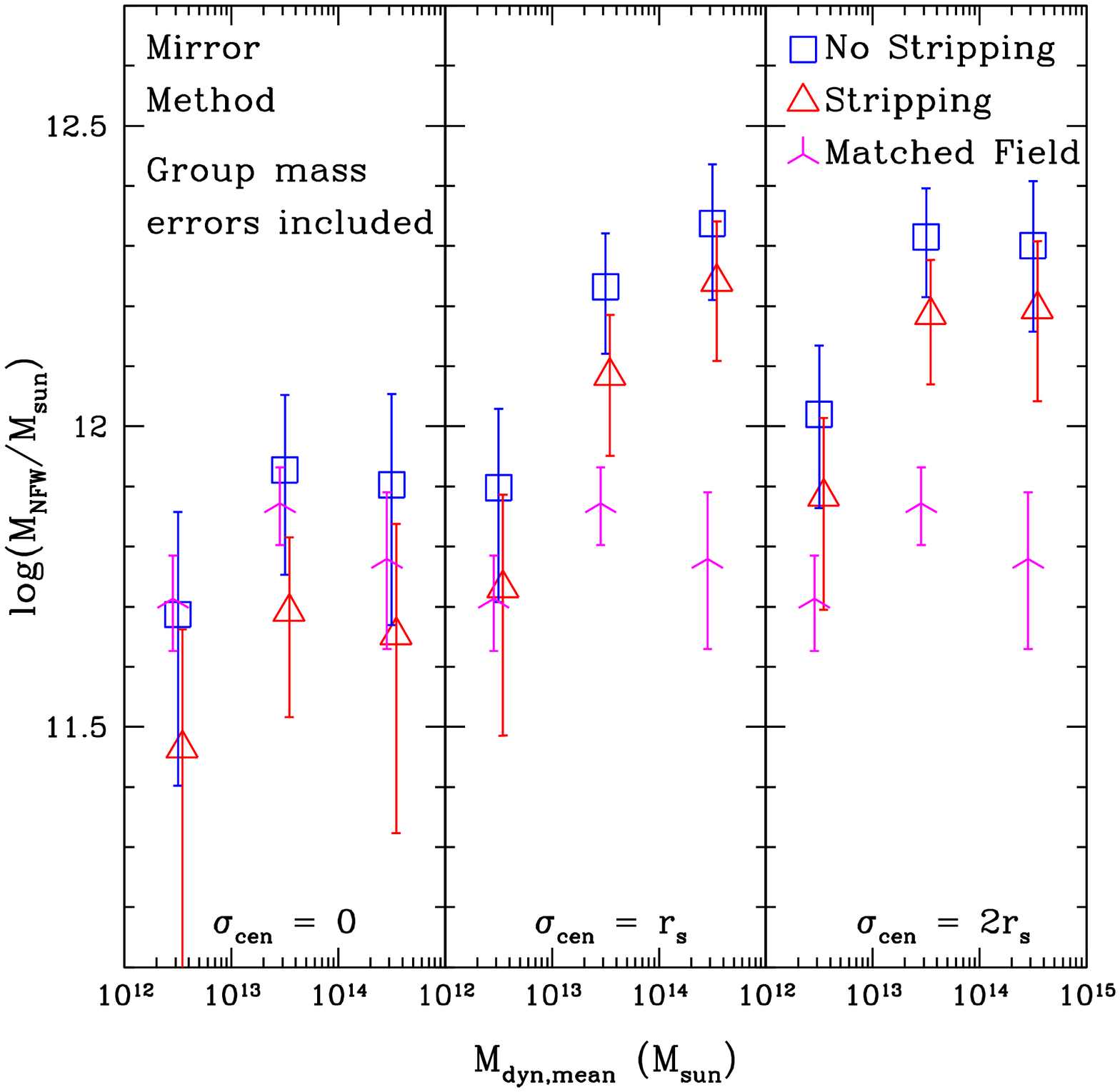}
  \includegraphics[scale=0.4]{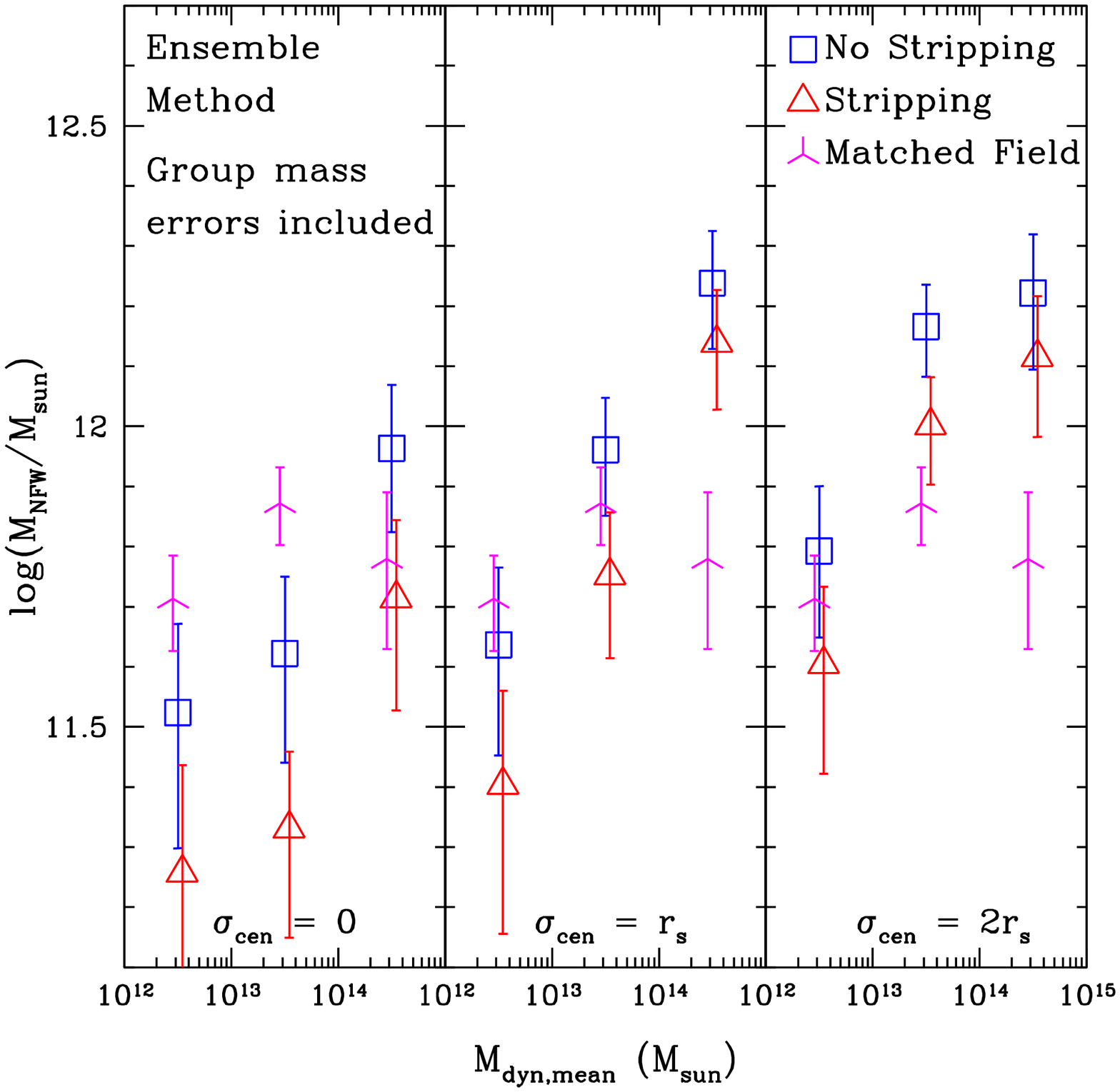}
  \hfill
  \caption[Comparison plot for the best-fit NFW mass with the Mirror Method for various group mass ranges and levels of centering error.]{Same as \figref{sc_mirrorcomp}, except satellites have been binned by their groups' estimated dynamical masses instead of the actual masses.}
  \label{fig:sc_mirrorcompme}
  \label{fig:sc_sigcompme}
\end{figure*}

Figure \ref{fig:sc_sigcompme} shows  the projected best-fit NFW masses for different group mass bins if there is significant error in our estimates of group masses. This causes bleeding between bins, increasing the expected mass in all group mass bins. We've illustrated this effect by using the dynamical mass for binning. Note that estimates of halo mass from stellar mass will likely have much lower scatter, so this should be seen as an upper bound for the effect of binning error.

\section{Discussion}
\label{sec:conc}

As our tests have shown, tidal stripping will have a significant effect on the lensing signals of satellites within groups. However, various sources of error make the detection of this effect more difficult in practice. As with any weak lensing measurements done, shape noise is the largest source of error and the most difficult to overcome -- it can only be reduced by gathering more data. By determining the expected signals for stripped and unstripped satellites, we can estimate the amount of data that will be necessary to discriminate between stripping models.

To measure tidal stripping, it's necessary to eliminate the contribution of the shared group halo to the lensing signal around satellites. Any method to do this will introduce its own sources of error, which can be comparable in magnitude to shape noise. In the cluster regime, it's possible to get very accurate mass estimates for individual clusters, and so these errors can be minimized even for individual clusters. In the group regime, it is only possible to estimate group mass sufficiently well when multiple groups are stacked together, but the far greater numbers of groups make this feasible.

\subsection{Optimal Methodology}

Our tests of both the Mirror Method and Ensemble Method show that in ideal circumstances, they both successfully eliminate the groups' contribution to the lensing signals around satellites. In this scenario, the Mirror Method is less biased, but suffers from a greater error due to shape noise - a result of the fact that it relies in the difference in signal between two points, causing the error to be increased by a factor of $\sqrt{2}$. The Ensemble Method causes errors to be contributed from the estimate of the average group mass, but the total errors still tend to be less than the errors from the Mirror Method.

When comparing the ``No Stripping'' and ``Stripping'' models, we see that the difference between the two models is larger when the Ensemble Method is used. This is due to the fact that the Mirror Method, by subtracting out the signal from around sample points, also subtracts out a portion of the lensing signal around satellites. In particular, for a point-mass satellite a distance $R$ from the centre of its group, the Mirror Method will detect zero signal for it on average for sources at annuli of radius greater than $2R$ (the distance from the satellite to the sample point). The satellites which are found closest to group centres tend to be the ones which have undergone the most stripping, but these are also the satellites which lose the largest amount of their own lensing signals when the Mirror Method is used. This suppresses the difference between the ``No Stripping'' and ``Stripping'' models.

Although neither method works perfectly when centering errors are present, the Ensemble Method shows somewhat less sensitivity to centering errors than does the Mirror Method. This difference is most significant for centering errors of $\sigma = r\sbr{s}$. Note that we expect to be able to quantify centering errors through \citet{GeoLeaBun12}'s method of fitting the lensing signal around group centres with an NFW profile convolved with centering error, but it is still useful to minimize the expected offset due to centering errors.

Overall, the Ensemble Method is preferable to the Mirror Method. It presents a better signal-to-noise in the difference between the expected signals for stripped and unstripped satellites, and it is less vulnerable to centering errors than the Mirror Method.

Other methods, such as a maximum-likelihood analysis, which work similarly to the Ensemble Method, will likely have a similar signal-to-noise to what we've calculated here.

\subsection{Prospects for Stripping Detections}

A detection of tidal stripping with either of the methods we have proposed will first require the issue of centering errors to be addressed. If they can be addressed and corrected for (eg. by using the method of \citet{GeoLeaBun12}), the Ensemble Method then provides the better S/N. By using the Ensemble Method and comparing the S/N in the difference between the Stripping and No Stripping models, we can estimate how much data will be needed in different group mass regimes in order to differentiate between the models. \tabref{tab1} lists our predicted S/N values for a measurement using the overlapping region of the CFHTLenS and GAMA-II surveys. We list S/N values for both the difference between the ``No Stripping'' and ``Stripping'' models, and the difference between the ``Matched Field'' dataset and the ``Stripping'' model. Of these two, the latter comparison is what we will measure in observational data, but it may be influenced by systematic errors in the application of the Ensemble Method. The former comparison is less likely to be influenced by systematic errors, as we used the same fields and realization of noise for both the ``No Stripping'' and ``Stripping'' models. As such, it provides a more realistic prediction for what we might see with observational data.

Of the three mass regimes tested, we see the highest S/N, $1.88$, in the group mass bin of $ 10^{13}$~M$_{\odot} < M\sbr{tot} < 10^{14}$~M$_{\odot} $. Although this doesn't quite reach the $2\sigma$ threshold for a detection, it is still worth making this measurement for various reasons:

\begin{itemize}
\item We do not apply any weighting scheme to our lenses or sources. An optimal weighting scheme could serve to boost the S/N.
\item When all three mass bins are combined together, the predicted S/N rises to $2.86$. This will allow us to detect the presence or absence of stripping, but not the group mass regimes in which it is taking place.
\item If stripping does occur and is stronger than predicted by our models (see \eqref{MretfromR}), a detection will be more likely.
\item Even if no $2\sigma$ detection can be made from the CFHTLens+GAMA-II dataset, the results of this measurement can be combined with future measurements to potentially add significance to those results.
\end{itemize}

\begin{table*}
\centering
\caption{Predicted detection S/N for measurements in various surveys, assuming the Ensemble Method is used, and using the comparison of our ``No Stripping'' and ``Stripping'' models.}
\begin{tabular}{|c c c c c|}
 & & \multicolumn{3}{c}{Predicted S/N$\sbr{NS-S}$} \\ \hline
$\log{M}$ & & CFHTLenS+GAMA-II & KIDS+GAMA-I & CFHTLenS+KIDS+GAMA \\ \hline
12--13 & & 1.43 & 1.63 & 2.17 \\
13--14 & & 1.88 & 2.14 & 2.85 \\
14--15 & & 1.61 & 1.84 & 2.44 \\
\hline
\end{tabular}
\label{tab:tab2}
\end{table*}

Our predicted significance can be improved upon if data from other surveys is combined with the CFHTLenS+GAMA-II dataset. Of note, KIDS \citep{VerKui12} will overlap the GAMA-I survey in 150 deg.$^2$ when it is complete, and we estimate it will contain an effective 8 sources per arcmin.$^2$. Assuming that the overlap between KIDS and GAMA-I will be $80\%$ unmasked, this results in a dataset that is $\sim1.3\times$ larger than CFHTLenS+GAMA-II. \tabref{tab2} shows the predicted significances for the three mass regimes we've tested, assuming the KIDS+GAMA-I dataset is used alone, and also combining it with the CFHTLenS+GAMA-II dataset. KIDS+GAMA-I alone is somewhat better than CFHTLenS+GAMA-II, and the combination of the two datasets has a $>50\%$ chance of a $2\sigma$ detection in each mass bin, if stripping is occurring at the strength we've modeled.

In conclusion, we will soon be able to use weak gravitational lensing to measure tidal stripping within even low-mass galaxy groups. A better understanding of tidal stripping will hopefully allow us insight into its role in the evolution of galaxies within the group environment.

\section{Conclusions}

In this paper, we have shown that for analysis of the lensing signals of satellite galaxies, the Ensemble Method (see \secref{Ensemble}) successfully subtracts out the contributions of these satellites' host groups to their lensing signal. This method performs better overall than the Mirror Method (see \secref{Mirror}).

We have shown here than errors in identifying the centres of group halos have a significant impact on the analysis of satellites' lensing signals. The contribution of these centering errors to the lensing signals around satellites is typically larger than the effect of tidal stripping on these signals. It is thus necessary to properly estimate the amount of centering error present in a sample before an assessment can be made of the presence or absence of tidal stripping.

If centering errors in the dataset can be measured, upcoming data from the GAMA-II survey, when combined with the CFHTLenS dataset, may be able to detect tidal stripping within groups of mass $ 10^{13}$~M$_{\odot} < M\sbr{tot} < 10^{14}$~M$_{\odot} $ to 2$\sigma$ significance, depending on the strength of stripping in reality and the scatter of the observational data. Data from KIDS and GAMA-I can be used to improve the significance of this measurement, and to also potentially detect stripping in lower-mass groups.

\section*{Acknowledgments}

We acknowledge useful discussions with Matt George, Catherine Heymans, Hendrik Hildebrandt, and Laura Parker. 

SH and JH acknowledge support by the DFG within the Priority Programme 1177 under the projects SCHN 342/6 and WH 6/3 and the Transregional Collaborative Research Centre TRR 33 ``The Dark Universe''. SH also acknowledges support by the National Science Foundation (NSF) grant number AST-0807458-002.

This work was made possible by the facilities of the Shared Hierarchical Academic Research Computing Network (SHARCNET:www.sharcnet.ca) and Compute/Calcul Canada.

\bibliographystyle{mn2e}
\bibliography{mjh}

\bsp

\appendix

\section{Approximations Used}

\subsection{Deriving Halo Mass from Stellar Mass}

\label{sec:MhfMs}

In \secref{Models}, we found it necessary to estimate the mass of a galaxy's dark matter halo, given its stellar mass. While \eqref{MhfromMs} can be easily used to estimate a galaxy's stellar mass if its halo mass is known, the equation is not invertible. We accomplish our task of estimating $M\sbr{halo}$ through an iterative process:

\begin{enumerate}
\item Begin with an estimate of the galaxy's halo mass, $M\sbr{init}$. For this, we use the $M\sbr{200}$ value for the galaxy given in \citet{DeBla07}'s database.
\item Set $M_{-1}$, $M_{-2}$, $M_{-3}$, and $M_{-4}$ all equal to $M\sbr{init}$.
\item Set $M\sbr{test}$ equal to $\left(M_{-1}+M_{-2}+M_{-3}+M_{-4}\right)/4$.
\item Determine $M\sbr{new}$ through:
\begin{eqnarray}
\label{eq:MhfromMs_iterate}
M\sbr{new} = \frac{M\sbr{stellar}}{0.129}\Biggl(\left(\frac{M\sbr{test}}{10^{11.4}\textrm{~M}_{\odot}}\right)^{-0.926}\nonumber \\
+\left(\frac{M\sbr{test}}{10^{11.4}\textrm{~M}_{\odot}}\right)^{0.261}\Biggr)^{2.44}\textrm{.}
\end{eqnarray}
\item If $M\sbr{new}$, $M_{-1}$, $M_{-2}$, $M_{-3}$, and $M_{-4}$ all differ by no more than 0.1\%, return $M\sbr{new}$ as $M\sbr{halo}$.
\item Set $M_{-4} = M_{-3}$, $M_{-3} = M_{-2}$, $M_{-2} = M_{-1}$, and $M_{-1} = M\sbr{new}$
\item Repeat steps 3 to 6 until $M$ has converged, for a maximum of 1,000 loops
\item If $M$ has not converged, return $M\sbr{init}$ as $M\sbr{halo}$
\end{enumerate}

The use of multiple past estimates of $M\sbr{halo}$ in our code was found to be necessary in order to ensure convergence, as when fewer past estimates were averaged and used, the algorithm would often end up alternating between two vastly different estimates for $M\sbr{halo}$. With the algorithm as written, $M\sbr{halo}$ was found to converge in every case tested in a sample 4$\times$4~deg.$^2$ field.

\subsection{Altering Concentration of NFW Profiles to Preserve Cusp Distribution}

\label{sec:cpfc}

In \secref{Models}, we mentioned an alteration we made to the NFW mass profile which leaves the density profile unchanged near the centre, but is lower than the initial profile at large radii. To illustrate how this works, examine the formula for the NFW density profile, with $\delta_c$, $r_s$, and $x$ expanded:

\begin{eqnarray}
\label{eq:NFWexp}
\rho(x) = \frac{ M\sbr{200} * (\ln(1+c)-c/(1+c))^{-1}}{4\pi \left(\frac{r\sbr{200}}{c}\right)^3}*\nonumber \\
\frac{1}{\left(\frac{rc}{r\sbr{200}}\right)\left(1+\left(\frac{rc}{r\sbr{200}}\right)\right)^2}\textrm{.}
\end{eqnarray}

In the limit $r \ll r\sbr{200}/c$, this becomes:

\begin{equation}
\label{eq:NFWexplim}
\rho(x) = \frac{M\sbr{200}}{4\pi r\sbr{200}^2}\frac{1}{r}\frac{c^2}{\ln(1+c)-c/(1+c)}\textrm{.}
\end{equation}

Since we modify the NFW profile by introducing a scale factor $f_r$ to the profile and modifying $c$, \eqref{cpfromcf} must be satisfied for the density in the core region to remain unchanged. As this equation is not solvable for $c'$, we use the following approximation: 

\begin{eqnarray}
\label{eq:cpapprox}
c' = 0.117 lc^4-1.70 lc^3+11.10 lc^2-30.7 lc+32.8\textrm{,}
\end{eqnarray}

where:

\begin{equation}
\label{eq:lcdef}
lc = \ln\left(c^2/\left(\left(\ln(1+c)-c/(1+c)\right)*f_r\right)\right)\textrm{.}
\end{equation}

\begin{figure}
\label{fig:cerr}
  \includegraphics[scale=0.4]{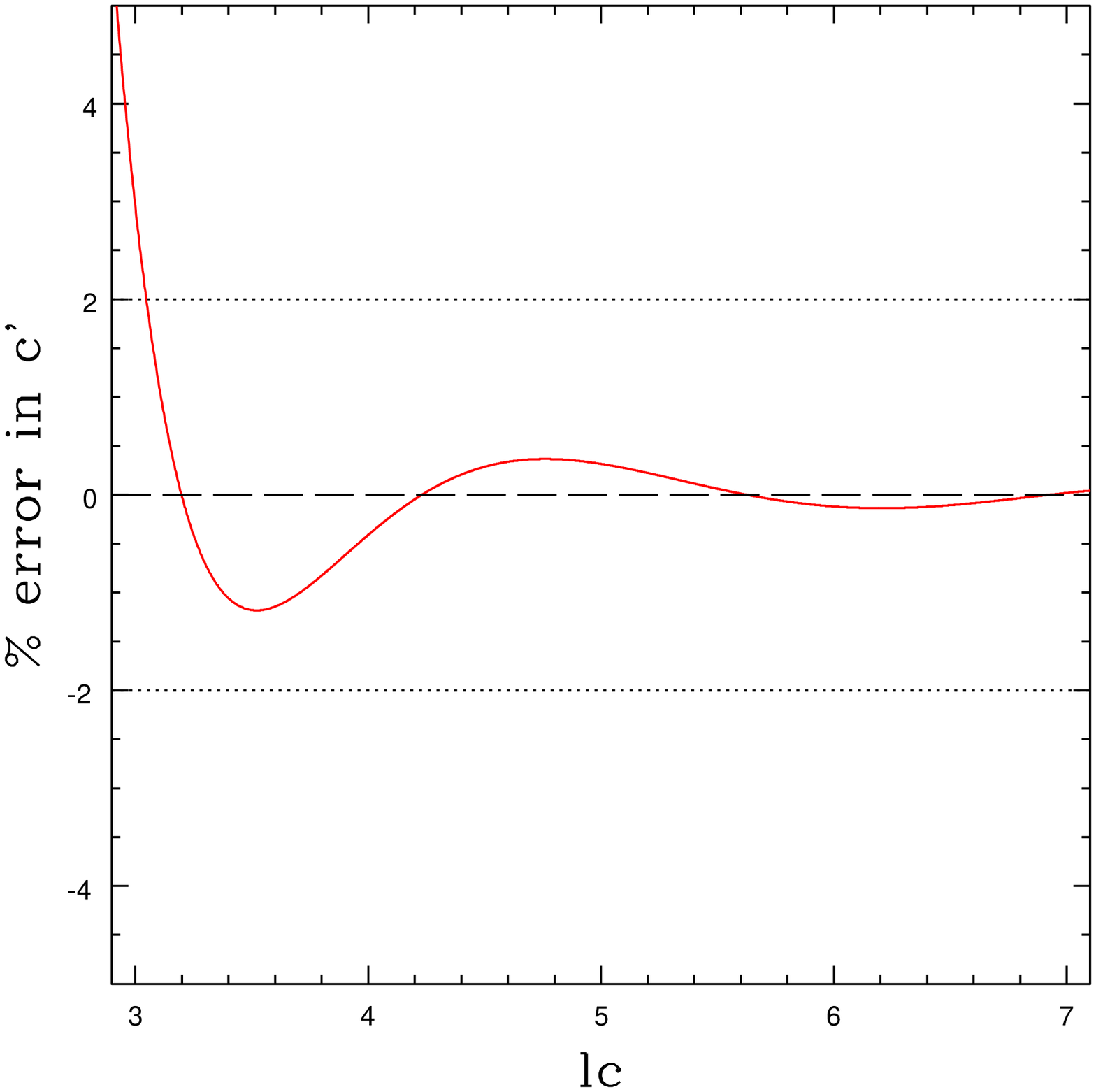}
  \caption[\% error in our estimates of c'.]{Percentage error in our estimates of $c'$, as calculated from \eqref{cpapprox}, as a function of lc (defined in \eqref{lcdef}). The error is less than 2\% for all reasonable values for the group's initial concentration $c$ and the fraction of mass $f_r$ retained by the group's halo. Significant error is only seen for extremely low concentration halos, which aren't modeled in our simulations.}
\end{figure}

This approximation has less than 2\% error for $3<lc<7$, as can be seen in \figref{cerr}, which covers the range of $lc$ possible for $4 < c < 15$ and $0.1 < f_r < 1$.
\label{lastpage}

\end{document}